\documentclass[aps,prc,twocolumn,superscriptaddress,nofootinbib,showpacs]{revtex4-1}
\usepackage{hyperref}
\usepackage{graphicx}
\usepackage{multirow}
\newcommand{\tabincell}[2]{\begin{tabular}{@{}#1@{}}#2\end{tabular}}
\usepackage{bm}
\begin{document}

\title{Production of light hypernuclei with light-ion beams and targets}

\author{Y.L.~Sun}
\email{yelei.sun@cea.fr}
\affiliation{D\'{e}partement de Physique Nucl\'{e}aire, IRFU, CEA, Universit\'{e} Paris-Saclay, F-91191 Gif-sur-Yvette, France\\}
\author{A.S.~Botvina}
\affiliation{Frankfurt Institute for Advanced Studies and ITP J. W. Goethe University, D-60438 Frankfurt am Main, Germany\\}
\affiliation{Institute for Nuclear Research, Russian Academy of Sciences, 117312 Moscow, Russia}
\author{A.~Obertelli}
\affiliation{D\'{e}partement de Physique Nucl\'{e}aire, IRFU, CEA, Universit\'{e} Paris-Saclay, F-91191 Gif-sur-Yvette, France\\}
\affiliation{Institut f\"{u}r Kernphysik, Technische Universit\"{a}t Darmstadt, D-64289 Darmstadt, Germany\\}
\author{A.~Corsi}
\affiliation{D\'{e}partement de Physique Nucl\'{e}aire, IRFU, CEA, Universit\'{e} Paris-Saclay, F-91191 Gif-sur-Yvette, France\\}
\author{M.~Bleicher}
\affiliation{Frankfurt Institute for Advanced Studies and ITP J. W. Goethe University, D-60438 Frankfurt am Main, Germany\\}

\date{\today}

\begin{abstract}
Ion-ion collisions at relativistic energies 
have been shown recently to be a promising technique for the production of hypernuclei.
In this article, we further investigate the production of light $\Lambda$ hypernuclei by use of a hybrid dynamical model,
cascade-coalescence followed by Fermi breakup.
The predictions
are then compared with the 
available experimental data.
The dependence of the production cross section upon the beam energy, beam mass number as well as different projectile-target combinations is investigated. 
In particular, we evaluate the yields and 
signal-over-background ratio
in the invariant-mass spectrum for carbon projectiles impinging on hydrogen and carbon targets
and various coincidence conditions in the experiment using the theoretical calculation as an input. 
It is found that 
comparing with carbon target, hydrogen target also leads to sizable hypernuclear yields, 
even for exotic species,
and the hydrogen target 
could improve significantly signal-over-background ratio in some hypernuclear invariant mass studies.
\end{abstract}


\maketitle

\section{INTRODUCTION}

Hyperons ($\Lambda$, $\Sigma$, $\Xi$, $\Omega$) are baryons containing at least one strange quark, unlike nucleons (proton or neutron) only composed of $u$ and $d$ valence quarks. 
The free $\Lambda$ particle, the lightest hyperon, can only decay to pion-nucleon system through the weak interaction, since the strong interaction conserves strangeness.
Interestingly, it was discovered that hyperons can form bound systems with nucleons and create short-lived hypernuclei \cite{Danysz1953}.
The investigation of hypernuclei provides
a practical method to study the fundamental hyperon-nucleon (YN) and hyperon-hyperon (YY) interactions
in nuclear matter at low energies.
Indeed, the very short lifetime of $\Lambda$ (263 ps \cite{Amsler2008}) makes it technically extremely difficult to 
use lambda particles directly as projectiles for scattering or capture experiments.

Starting from the end of 60's, emulsion technique with cosmic rays 
and missing mass technique using 
$^A$Z(K$^-$, $\pi$$^-$)$^A_\Lambda$Z and $^A$Z($\pi$$^+$, K$^+$)$^A_\Lambda$Z reactions 
have been widely used to produce hypernuclei in the laboratory \cite{Gal2016}.
As the lambda particle can be distinguished from nucleons by the strangeness quantum number and is not limited by the Pauli exclusion principle \cite{Hotchi2001},
adding a lambda particle
to a nucleus tends to increase the binding energies of the whole system.
The measured binding energies 
thus give an information on the YN interaction.
Moreover, the glue-like role of the hyperon is expected to change the nuclear deformation \cite{Tanida2001, Hagino2013}, 
to lead to new excitation modes \cite{Hagino2013}, 
and to shift the neutron and proton drip line from their normal limits \cite{Samanta2008, Botvina2013}. 
They have been discovered to bind loosely unbound nuclei 
at the drip line such as $^{5}$H and $^{7}$He (via $^6_{\Lambda}$H \cite{Agnello2012} and $^{8}_{\Lambda}$He \cite{Juric1971}, respectively).
Hyperons are also expected to be of major interest for nuclear structure: due to the absence of Pauli blocking with nucleons, 
they constitute 
a unique opportunity to probe the inner densities in nuclei.
Furthermore, in the field of astrophysics, hyperons are predicted to exist inside neutron stars at
densities exceeding 2-3$\rho$$_{0}$, where $\rho$$_{0}$ = 0.16 baryon/fm$^3$, which is the nuclear saturation density.
However, depending on the detailed properties of the YY interaction and YNN three-body interaction, 
the presence of hyperons in neutron stars can either soften or stiffen the high-density equation of state (EOS), 
resulting in large 
uncertainty
in the prediction of the maximum mass of neutron stars \cite{Zdunik2013, Lonardoni2015}. 
The study of the neutron-rich hypernuclei may provide relevant information to solve this hyperon puzzle 
and help us to achieve a better description on the EOS of 
high density nuclear matter and the evolution of compact stars.

Although many theoretical works predict the existence of neutron-rich or proton-rich hypernuclei, however, 
up to now most of the produced hypernuclei are limited to 
systems close to the stability line.
Very recently, two distinguished results were obtained using the double charge exchange reaction 
$^6$Li(K$^-$$\rm_{stop}$, $\pi$$^+$)$^6_\Lambda$H \cite{Agnello2012} and 
the electroproduction reaction $^7$Li($\emph{e}$, 
$\emph{e}$${^\prime}$K$^+$)$^7_\Lambda$He \cite{Nakamura2013},
with very small cross sections ($\approx$10 nb/sr). 
The production of heavier exotic hypernuclei with large isospin asymmetry using the techniques mentioned above is not feasible.
Reactions with ion beams at relativistic energies provide an alternative approach to 
overcome this bottleneck. 
Searching for hypernuclei using ion beams can be traced back to 70's at Berkeley \cite{Nield1976} and later in Dubna \cite{Avramenko1992}, 
although at that time, the produced hypernuclei were only signaled 
by hardware trigger selection 
without detail particle identification.
With the improvement of the experimental setup, $^3_\Lambda$H and $^4_\Lambda$H were successfully identified in central collisions 
by impinging 11.5 GeV/c platinum beams on a Au target \cite{Armstrong2004}. 
Recently, 
hypertriton ($^3_\Lambda$H) and antihypertriton ($^3_{\bar{\Lambda}}$$\bar{\rm{H}}$)
were observed by the STAR collaboration in the colliding of two Au beams at an energy of 200 GeV/nucleon at RHIC\cite{Star2010},
as well as by the ALICE collaboration in ultra-relativistic collisions at LHC \cite{Alice2013}.
It is considered that in such central ion collisions
only light hyper clusters with A$\leq$4 could be produced because of the very high temperature of the fireball ($\emph{T}$ $\approx$ 160 MeV) \cite{Andronic2011}. 
However, the coalescent mechanism of cluster formation may lead to formation of more heavy species outside the mid-rapidity region \cite{Botvina2015}.

More recently, 
the known hypernuclei $^3_\Lambda$H and $^4_\Lambda$H, with their lifetimes, were measured 
in the projectile rapidity region by
the HypHI collaboration via fragmentation of a 2\emph{A} GeV $^6$Li beam impinging on a carbon target \cite{Rappold2013NPA, Rappold2013PRC}. 
The experiment successfully demonstrates the feasibility of producing hypernuclei in the peripheral collisions. 
Note that in this technique, large fragments of projectile and target nuclei do not interact with each other intensively and 
form spectator residues, which might capture the hyperons if momentum matching allows.
Due to the large Lorentz boost, the production and decay vertices can be well separated by tens of centimeters, 
making it possible to identify
effectively
the production and decay of hypernuclei independently. 
In addition, a possible existence of $^3_\Lambda$\emph{n} was suggested, which might come from the 
disintegration
of heavier projectile-like 
hyperfragments \cite{Rappold2013PRC}, implying a new mechanism of producing exotic hypernuclei. 
In the near future, the FAIR \cite{Fair2006, Aysto2016} facility in GSI and HIAF \cite{Yang2013} facility in China will provide high energy and high intensity ion beams,
providing good opportunities to study projectile-like hypernuclei.
Proton-rich and neutron-rich hypernuclei are foreseen to be produced efficiently using various secondary beams 
\cite{Botvina2013, Rappold2016}.

In the present article, we focus on the production of hypernuclei in peripheral ion collisions.
The production of hypernuclei is investigated in the whole rapidity region by 
considering the cascade-coalescence as well as the projectile and target decay processes.
We first give a description of our model in section 2 and then compare the calculation results with few existing data in section 3.
We then present new calculations to investigate the energy dependence, the projectile 
and target dependence 
to produce light $\Lambda$ hypernuclei in section 4.
Finally, in section 5, we comment on the signal-over-background ratio as 
a function of target-projectile combinations and particle coincidences in possible future experiments.

\section{MODEL DESCRIPTION}
In high-energy ion collisions, 
the main production sources of hyperons are nucleon-nucleon collisions, \emph{e.g.} \emph{p}+\emph{p}$\rightarrow$ \emph{p}+$\Lambda$+K$^+$(threshold $\emph{E}_{lab}$ $\geq$ 1.58 GeV) 
and also secondary meson-nucleon collisions, $\emph{e.g.}$ $\pi^+$+$\emph{n}$$\rightarrow$$\Lambda$+K$^{+}$(threshold $\emph{E}_{lab}$ $\geq$ 0.76 GeV). 
At energies lower than 2 GeV/nucleon, these elementary collisions are reliably described by using the available experimental data and phenomenological parameterizations.
Experimental cross sections, or calculated cross sections if data is not available, are used to calculate the angular and energy distributions of outgoing channels.
At higher energies, the formation of hyperons is usually estimated 
by hadronization models,
such as the quark gluon string model (QGSM) \cite{Toneev1990}, PYTHIA \cite{Torbjorn2006, Torbjorn2015} or 
the Lund FRITIOF string model \cite{Nilsson1987}.
Sub-hadronic degrees of freedom, such as quarks and strings, are taken into account phenomenologically.
Afterwards, the evolution of hadrons (mesons + baryons) in space and time can be described by 
transport models ($\emph{e.g.}$, DCM \cite{Toneev1983, Toneev1990}, UrQMD \cite{Bass1998, Bleicher1999}, GiBUU \cite{Buss2012} and HSD \cite{Geiss1998, Cassing2008})
by solving the relativistic Boltzmann transport equations.
All these models have been shown to be successful in the description of hyperon production \cite{Botvina2011, Gaitanos2009, Botvina2015}.

During the transportation, the capture of lambda particles and other particles 
by the neighboring nuclear fragments is determined 
either by the coalescence criterion \cite{Wakai1988, Steinheimer2012, Botvina2015}, 
$\emph{i.e.}$ hypernuclei are formed when hyperons are close to
nucleons
in both the spatial and momentum space, 
or by the potential criterion \cite{Botvina2011}, 
$\emph{i.e.}$ hypernuclei are formed when the kinetic energies of hyperons in the residue-at-rest frame are smaller than the attractive potential of the residues. 
Since the primarily produced $\Lambda$ particles 
have usually large momentum mismatch with the fragments and thus hardly to be captured,
re-scattering and secondary interactions are important for the capture process \cite{Botvina2016EPJA}. 
A comparison on the absorption rate between DCM and UrQMD could be found in Ref. \cite{Botvina2011}, where qualitative agreement was reported. 

   \begin{figure}[t!]
   \setlength{\abovecaptionskip}{0pt}
   \setlength{\belowcaptionskip}{0pt}
   \begin{center}
   \includegraphics[width=8cm]{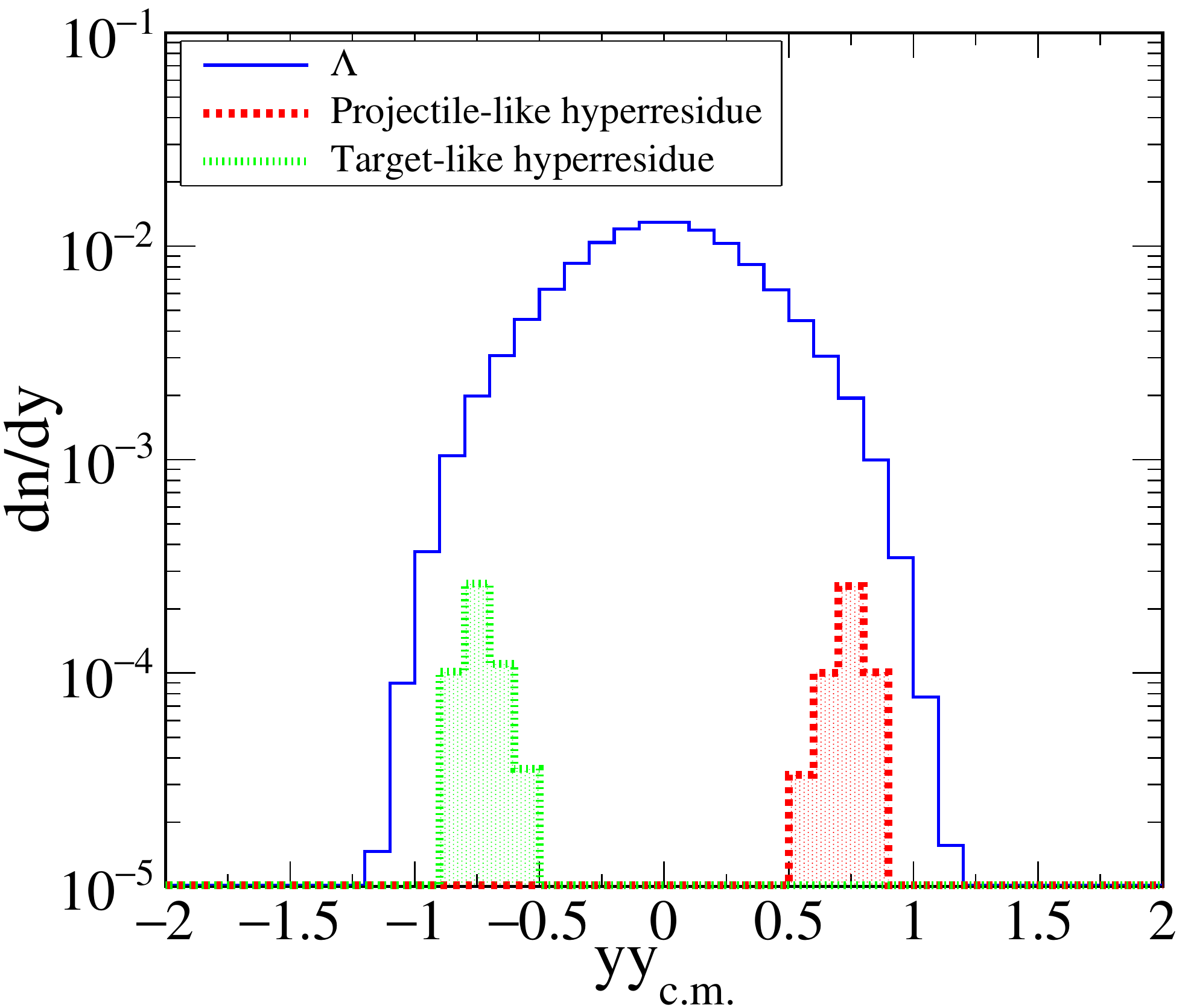}
   \end{center}
   \caption{(Color online) Rapidity distributions of $\Lambda$ hyperons and hyperresidues 
in the center-of-mass frame
as calculated in the DCM model for $^{12}$C (2\emph{A} GeV) + $^{12}$C collision. 
Solid blue line gives the rapidity of $\Lambda$ hyperons.
Red and Green histograms show the rapidity of projectile-like and target-like hyperresidues, respectively.}
   \label{fig:1}
   \end{figure}
   \begin{figure}[t!]
   \setlength{\abovecaptionskip}{0pt}
   \setlength{\belowcaptionskip}{0pt}
   \begin{center}
   \includegraphics[width=8cm]{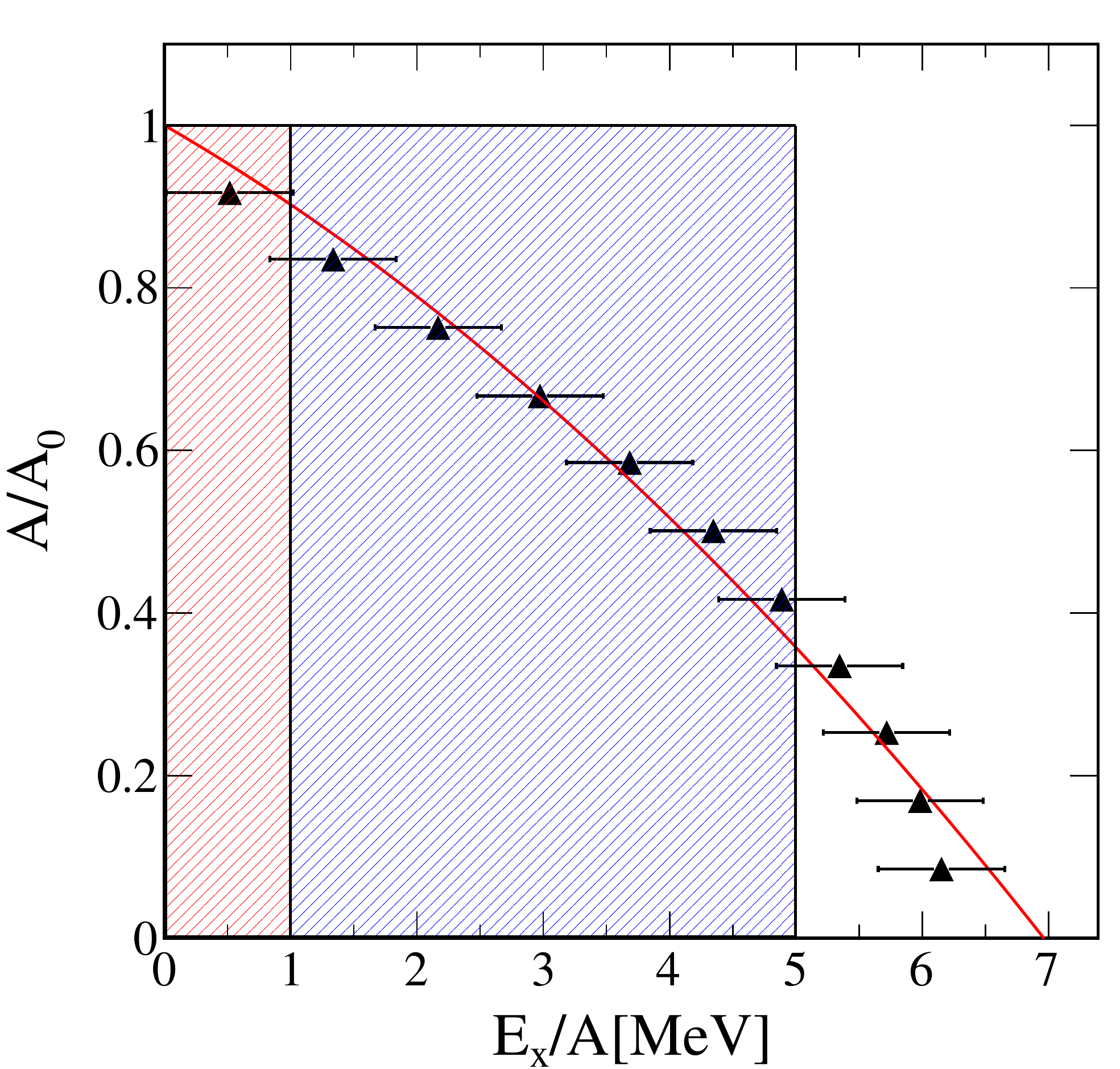}
   \end{center}
   \caption{(Color online) Excitation-energy distributions used in the present theoretical calculation. 
The red curve shows parameterized excitation-energy distribution of equation (\ref{equation:1}). 
The red and blue shadow areas show the other two uniform excitation-energy distributions adopted in the calculation.
The black points are taken from Ref. \cite{Botvina2017}, which are the excitation energies extracted using DCM.}
   \label{fig:2}
   \end{figure}
   \begin{figure*}[t!]
   \setlength{\abovecaptionskip}{0pt}
   \setlength{\belowcaptionskip}{0pt}
   \centering
   \includegraphics[width=15cm]{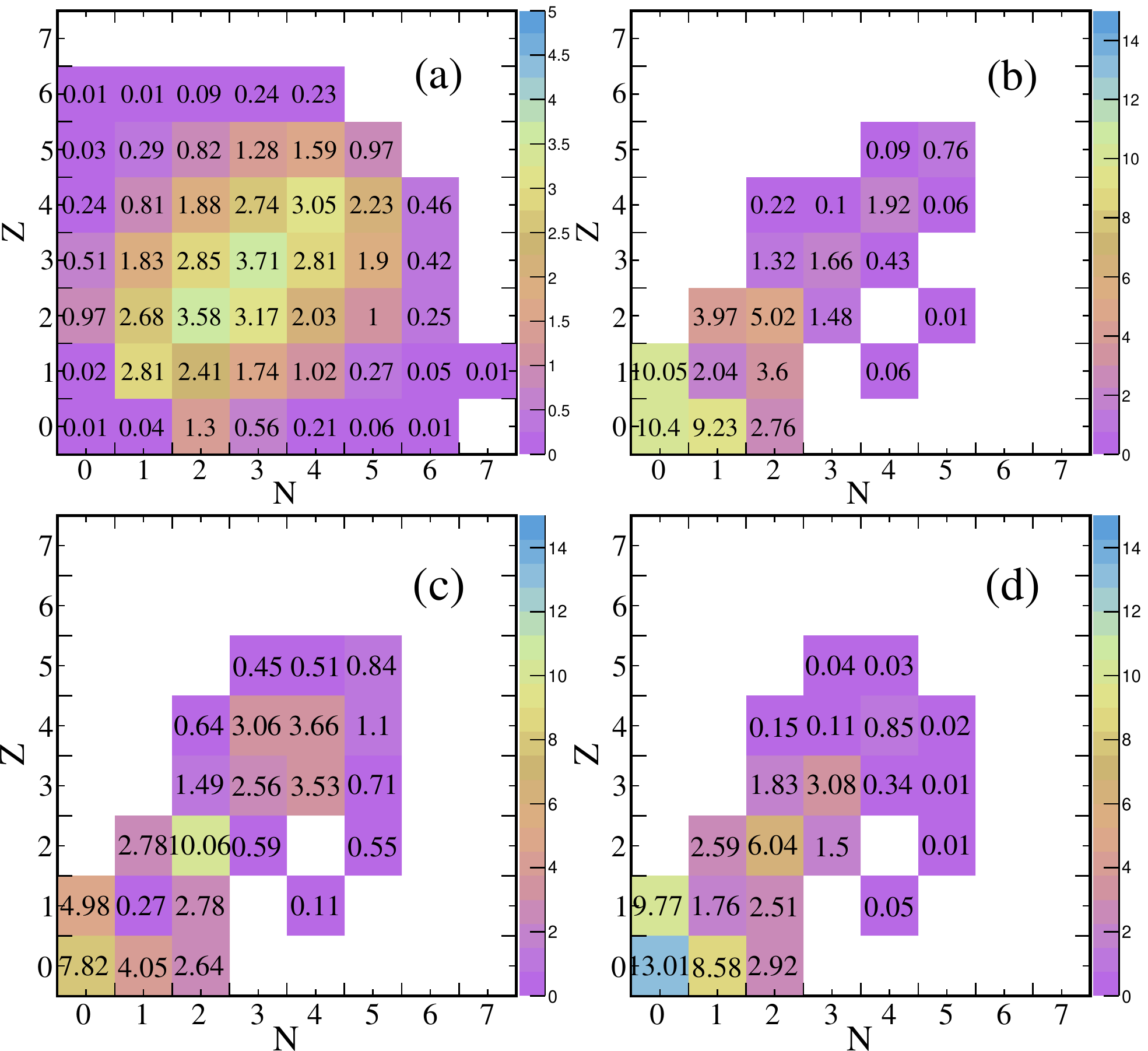}\\
   \caption{(Color online) Hypernuclear production cross sections for $^{12}$C + $^{12}$C at 2\emph{A} GeV. (a) without Fermi breakup, (b) with Fermi breakup and parameterized $\emph{E}_{x}$ distribution, (c) with Fermi breakup and $\emph{E}_{x}$: 0-1\emph{A} MeV uniform distribution, (d) with Fermi breakup and $\emph{E}_{x}$: 1\emph{A}-5\emph{A} MeV uniform distribution. The unit of the cross sections are given in $\mu$b.
Only projectile-like species are considered.
}
   \label{fig:3}
   \end{figure*}

After the non-equilibrium and absorption stages ($\tau \approx 50-100$ fm/c), 
as expected for normal residual nuclei \cite{Botvina1992, Hudan2003},
the formed primary hyperresidues are usually in high excited states.
They may de-excite with production of cold normal nuclei and hypernuclei \cite{Botvina2007},
which could take time ranging from 10$^{2}$ to 10$^{4}$ fm/c \cite{Botvina2011}.
Various de-excitation processes can take place depending on the excitation energy and the mass number of the hyperresidues.
For light excited hyperresidues (\emph{A} $\leq$ 16), Fermi-breakup model can be applied \cite{Lorente2011, Botvina2013}.
While for larger species, 
the evaporation-fission model ($\emph{E}$$^*$$\textless$ 2-3 MeV/nucleon) \cite{Botvina2016PRC},
or multi-fragmentation model \cite{Botvina2007} at higher excitation energy can be used.

In the present work we adopted the Dubna intranuclear Cascade Model (DCM) \cite{Toneev1983, Botvina2013}
for description of collisions with relatively light nuclei (\emph{A} $\leq$ 16).
For elementary hadron collisions at relatively low energy (\emph{E}$\rm_{lab}$ \textless 1-3 GeV/nucleon, where \emph{E}$\rm_{lab}$ is 
the laboratory energy of the colliding nuclei)
the model uses quite reliable approximations for the reaction channels supported by the analysis of large amount of available experimental data \cite{Botvina2017}.
In addition, at higher energies (\emph{E}$\rm_{lab}$ \textgreater 4.5 GeV/nucleon), the QGSM \cite{Toneev1990, Botvina2011} is involved with the smooth transition between these two limits \cite{Botvina2017}. 
For the absorption we took the potential criterion.
The hyperon potential in cold nuclear matter at the saturation density was taken as 30 MeV, and the correction on the density variation was applied \cite{Botvina2011}.
The de-excitation was described by the Fermi-breakup model \cite{Lorente2011, Botvina2013}.
In this model we have considered the decay channels of hot primary hyperresidues into all existing hypernuclei
(\emph{A} $\leq$ 16):
$^{3,4}_{\Lambda}$H,
$^{4,5,6}_{\Lambda}$He,
$^{6,7,8,9}_{\Lambda}$Li,
$^{7,8,9,10}_{\Lambda}$Be,
$^{10,11,12}_{\Lambda}$B,
$^{12,13,14}_{\Lambda}$C,
$^{14,15}_{\Lambda}$N,
$^{16}_{\Lambda}$O.
Some neutron-rich hypernuclei,
such as $^3_{\Lambda}$\emph{n}, $^6_{\Lambda}$H and $^{8}_{\Lambda}$He, were included
so that we could examine the production yields of exotic species.
Also, for complete analysis, exotic $^{2}_{\Lambda}$H and $^{2}_{\Lambda}$\emph{n} hypersystems were considered 
with masses equal to the sum of the masses of constituents. 
If there are no such bound states in reality, these species should be taken as unbound particles.

In Fig. \ref{fig:1}, we give an example for $^{12}$C + $^{12}$C collision at 2\emph{A} GeV. 
The rapidity of $\Lambda$, projectile-like and target-like hypernuclei in the center-of-mass frame are shown in different colors.
Yield of each component was normalized by the total number of the inelastic collisions.
From Fig. \ref{fig:1}, we could see that the produced lambda hyperons have a very broad rapidity distribution.
The broadening is mainly due to the $\Lambda$N elastic scattering whose typical cross section is around 30 mb \cite{Botvina2011}. 
Projectile-like and target-like hypernuclei can be formed in the overlap rapidity region
between the residues and the lambda hyperons.

In the de-excitation processes of the calculation, the excitation energy of the primary hypernuclei is determined afterward by using
a parameterized excitation-energy distribution, as shown in Fig. \ref{fig:2}.
The excitation-energy distribution can be described by the following equation, see \cite{Botvina1995},

\begin{equation}\label{equation:1}
A/A_{0} = 1 - a_{1}(E_{x}/A)-a_{2}(E_{x}/A)^{2}
\end{equation}

\noindent
where $\emph{a}$$_{1}$ = 0.08983 MeV$^{-1}$, $\emph{a}$$_{2}$ = 0.007728 MeV$^{-2}$,
\noindent
\emph{A} is the mass number of the hyperresidue and $\emph{A}_{0}$ is the mass number of the projectile or the target.
This correlation is consistent with
the result obtained in the DCM calculations
for light collision system ($^{12}$C + $^{12}$C at 5\emph{A} GeV) \cite{Botvina2017},
which is demonstrated by the black symbols in Fig. 2.
To further evaluate the sensitivity of the hypernuclear yields to the excitation energy of the primary hyperresidues,
we explore two other excitation-energy distributions.
$\emph{E}_{x}$ has uniform distribution from 0 to 1\emph{A} MeV and from 1\emph{A} to 5\emph{A} MeV, 
corresponding to red and blue shadow areas in Fig. \ref{fig:2}. 
The production cross sections of hyperfragments 
from $^{12}$C + $^{12}$C at 2 GeV/nucleon is shown in Fig.\ref{fig:3}. 
Panel (a) shows the production cross sections for all projectile-like residues after the cascade but before de-excitation 
while panel (b) shows the production cross sections after de-excitation following the parameterization of Eq. (\ref{equation:1}).
Panel (c) and (d) show the production cross sections after the de-excitation with excitation energy uniformly distributed from 0 to 1\emph{A} MeV
and from 1\emph{A} to 5\emph{A} MeV.
As shown in Fig. \ref{fig:3}(a), 
without de-excitation, 
just after the cascade-coalescence stage,
many hot primary hyperresidues could be produced. 
However, after considering the de-excitation, as shown in Fig. \ref{fig:3} (b), (c) and (d), 
most of the remaining cold hypernuclei locate close to the $\beta$-stability line with yields around several microbarns.
Neutron-rich or proton-rich hypernuclei could also survive with relatively smaller probability. 
Cross sections of $^6_{\Lambda}$H and $^{8}_{\Lambda}$He are on the order of few to several tens of nanobarns.
Note that $^{5}_\Lambda$H and $^{7}_\Lambda$He were not considered as bound systems at the breakup stage, 
which results in empty holes in the \emph{Z}-\emph{N} plane.
The hypernuclear yields
in the present work are smaller than those in Ref. \cite{Rappold2016},
and the form of $\emph{Z}$-$\emph{N}$ plane population is different.
The differences are due to that in Ref. \cite{Rappold2016} 
the geometrical cross sections instead of the inelastic cross sections were wrongly taken to estimate the hypernuclear yields,
and breakup of the hot primary hyperresidues was not considered
\cite{Rappold2016_private}.
Actually, the production cross sections are quite sensitive to the adopted excitation energies, since the breakup probability is proportional to ($E_{x}-U)^{3n/2-5/2}$, 
where $\emph{U}$ is the Coulomb barrier and $\emph{n}$ is the number of cold residues after breakup \cite{Lorente2011}.
Especially for heavier hyper-fragments, the resulting cross sections could be more than one order of magnitude of difference.
If no special indication is given,
the following calculations are performed with the parameterization of Eq. (1) for the excitation-energy function.

\section{Benchmark with existing data}
\begin{figure*}[t!]
   \setlength{\abovecaptionskip}{0pt}
   \setlength{\belowcaptionskip}{0pt}
   \centering
   \includegraphics[width=18.5cm]{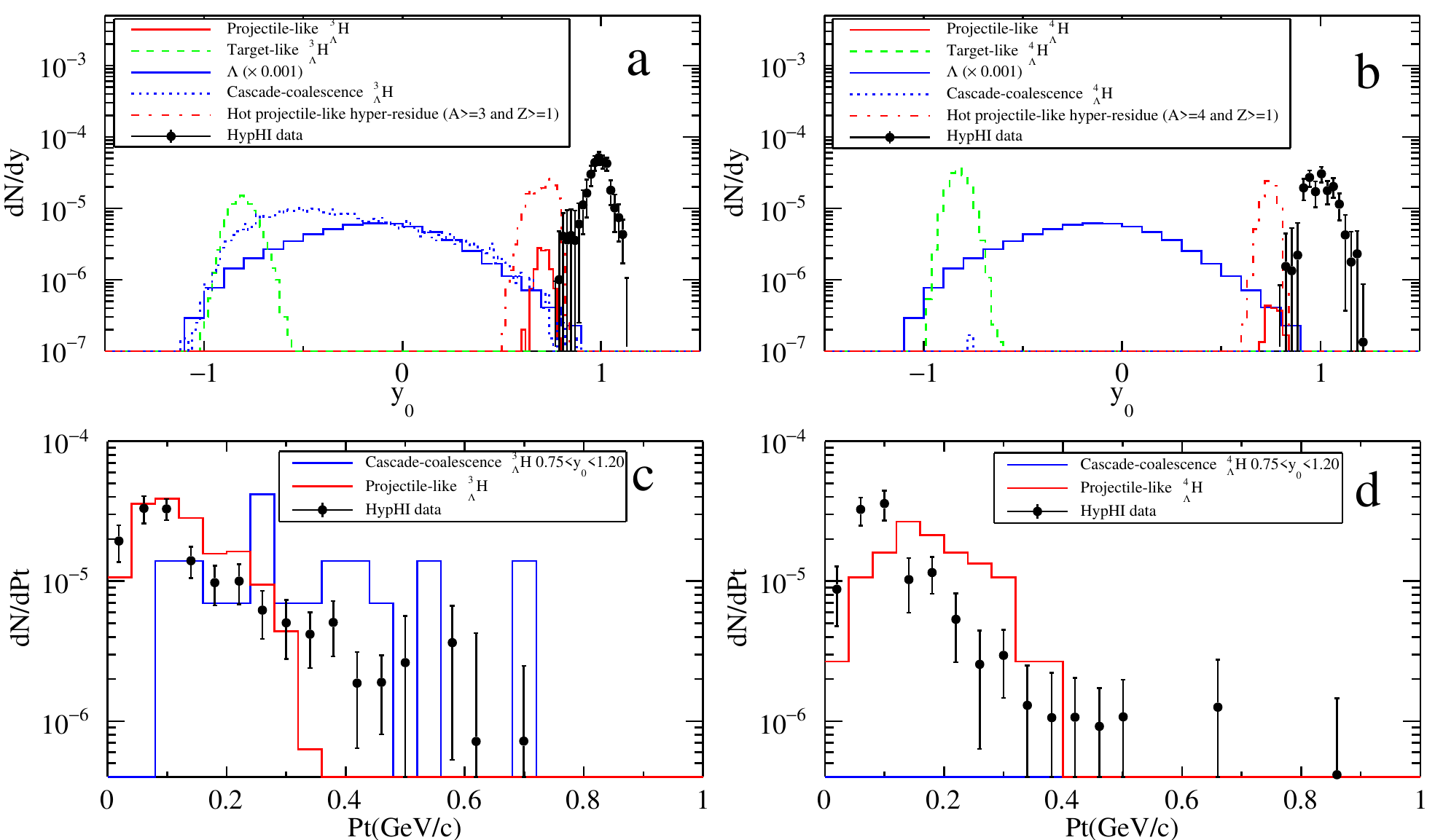}\\
   \caption{(Color online) Rapidity and transverse momentum distributions are compared with the experimental data of HypHI. 
y$_{0}$ denotes the rapidity in the center-of-mass frame of the individual NN collisions scaled to the rapidity of this reference frame
and P$_{t}$ is the transverse momentum.
Panel (a) and (b) show the rapidity results related to $^{3}_{\Lambda}$H and $^{4}_{\Lambda}$H. 
The projectile-like, target-like and cascade-coalescence hypernuclei are shown in different colors. 
The experimental data from the HypHI experiment \cite{Rappold2015} are shown in black points. 
Data is renormalized by divide the rapidity bin size of 
0.02 for $^{3}_{\Lambda}$H or 0.03 for $^{4}_{\Lambda}$H (See Fig. 3 of Ref. \cite{Rappold2015}).
The rapidity distributions are normalized by the total number of the inelastic collisions.
dN/dy is therefore the multiplicity per inelastic collision per unit of rapidity.
Panel (c) and (d) show the transverse momentum of $^{3}_{\Lambda}$H and $^{4}_{\Lambda}$H. 
The projectile-like and cascade-coalescence contribution in the forward rapidity region are shown in different colors. 
Data is renormalized by divide the momentum bin size of 40 MeV/c (See Fig. 3 of Ref. \cite{Rappold2015}).
The theoretical transverse momentum distributions are normalized by the data.
}
\label{fig:hyphi_rapidity}
\end{figure*}

Until now, the experimental cross sections of ion induced hypernuclear production are very scarce.
Two available data were measured at Dubna \cite{Avramenko1992} and at GSI by the HypHI collaboration \cite{Rappold2015}.
Both measurements use ion beams impinging on carbon target and focused on the formed hypernuclei around the beam rapidity region. 
The theoretical projectile-like hypernuclear yields 
and the experimental data
are summarized in Table \ref{table:1}.
\begin{table}
\caption{\label{table:1}
The calculated cross sections are compared with the Dubna and HypHI data.
Three excitation-energy distributions are considered in the calculation. 
(\uppercase\expandafter{\romannumeral1}) parameterized $\emph{E}_{x}$ distribution.
(\uppercase\expandafter{\romannumeral2}) and (\uppercase\expandafter{\romannumeral3}) 
$\emph{E}_{x}$ uniformly distributed from 0 to 1\emph{A} MeV and 1\emph{A} to 5\emph{A} MeV, 
considered as test excitation-energy distributions without theoretical foundations.
The unit of the cross sections are given in $\mu$b.
Only projectile-like hypernuclei are considered.
}
\begin{tabular*}{8.5cm}{@{\extracolsep{\fill}}c|cccc}
\hline
\hline
Beam & \tabincell{c}{Energy\\(GeV/nucleon)} &   &  $^3_\Lambda$H & $^4_\Lambda$H\\

\hline
\multirow{4}{*}{$^{3}$He} & \multirow{4}{*}{5.14}& (\uppercase\expandafter{\romannumeral1})  & 0.63      & \\
                          &                      & (\uppercase\expandafter{\romannumeral2})  & 0.05      & \\
                          &                      & (\uppercase\expandafter{\romannumeral3})   & $\textless$ 0.01  &  \\
                          &                      & Dubna \cite{Avramenko1992} & 0.05$^{+0.05}_{-0.02}$ & \\
\hline
\multirow{4}{*}{$^{4}$He} & \multirow{4}{*}{3.7} & (\uppercase\expandafter{\romannumeral1})   & $\textless$ 0.01        & 0.19 \\
                          &                      & (\uppercase\expandafter{\romannumeral2})   & 0.24      & 0.12         \\
                          &                      & (\uppercase\expandafter{\romannumeral3})   & 0.04      & $\textless$ 0.01   \\
                          &                      & Dubna \cite{Avramenko1992} & $\textless$ 0.1  & 0.4$^{+0.4}_{-0.2}$ \\
\hline
\multirow{4}{*}{$^{6}$Li} & \multirow{4}{*}{3.7} & (\uppercase\expandafter{\romannumeral1})  & 1.15 & 0.27 \\
                          &                      & (\uppercase\expandafter{\romannumeral2})  & 0.29 & 2.31 \\
                          &                      & (\uppercase\expandafter{\romannumeral3})  & 0.84 & 0.33 \\
                          &                      & Dubna \cite{Avramenko1992} &0.2$^{+0.3}_{-0.15}$ & 0.3$^{+0.3}_{-0.15}$ \\
\hline
\multirow{4}{*}{$^{6}$Li} & \multirow{4}{*}{2.0} &(\uppercase\expandafter{\romannumeral1})   & 0.2        & 0.02\\
                          &                      &(\uppercase\expandafter{\romannumeral2})   & 0.03       & 0.43\\
                          &                      &(\uppercase\expandafter{\romannumeral3})   & 0.13       & 0.04\\
                          &                      & HypHI \cite{Rappold2015}  & 3.9$\pm$1.4 & 3.1$\pm$1.0\\
\hline
\hline
\end{tabular*}
\end{table}
Here, we note that in the case of ($^{3}$He, $^{3}_\Lambda$H) and ($^{4}$He, $^{4}_\Lambda$H), 
one proton in the projectile is substituted by a lambda particle.
In the calculation using parameterized excitation-energy distribution, the yields of $^{3}_\Lambda$H and $^{4}_\Lambda$H are not reduced by the decay processes,
since the formed hyperresidues have the same mass number as the projectiles,
and the excitation energies deduced from equation (\ref{equation:1}) are always zero.
The model with such excitation-energy distribution thus 
is useful only for reactions of deep disintegrations of large projectiles.
Instead, the other two excitation-energy distributions (0 \textless $\emph{E}_{x}$/\emph{A} \textless 1 MeV or 1 MeV \textless $\emph{E}_{x}/\emph{A} \textless $ 5 MeV) reach a better agreement with the Dubna data.
In the case of $^{6}$Li projectile at 3.7 GeV/nucleon, the parameterized excitation-energy distribution reproduces the yield of $^{4}_\Lambda$H,
but overestimates the yield of $^{3}_\Lambda$H, while a smaller excitation energy (0 \textless $\emph{E}_{x}$/\emph{A} \textless 1 MeV) gives a better result.
The excitation-energy distribution thus seems to have a
more complicated dependence on the mass ratio (\emph{A}/$\emph{A}_{0}$) than that in equation (\ref{equation:1})
for lightest projectiles,
which is difficult to be benchmarked at this stage with limited data set.
Nevertheless, 
we conclude that a satisfactory agreement is found with the Dubna data.
On the other hand, we found the calculated yields of $^{3}_\Lambda$H and $^{4}_\Lambda$H in $^{6}$Li + $^{12}$C collisions at 2 GeV/nucleon
are more than one order of magnitude smaller than the HypHI data.
Surprisingly, we found none of the excitation distributions could result in same order of magnitude of cross sections as the HypHI experiment.
To investigate the reasons, we further compare the rapidity and transverse momentum distributions with the experimental data.
Note that here we can do direct comparision since the experimental rapidity and transverse momentum distributions 
have already been corrected by the experimental acceptance and reconstruction efficiency \cite{Rappold2015}.
The results are shown in Fig. \ref{fig:hyphi_rapidity}.
In panel (a) and (b), the distributions are plotted as a function of the particles' rapidity in the 
center-of-mass reference frame of the individual NN collisions which is scaled to the rapidity of this reference frame:
y$_{0}$ = (y$\rm_{lab}$-y$\rm_{cm}$)/y$\rm_{cm}$, 
where y$\rm_{cm}$ denotes the rapidity of the center-of-mass reference frame of the individual NN collisions.
\footnote{
For the collision of $^{6}$Li + $^{12}$C at 2$\emph{A}$ GeV,
y$\rm_{cm}$ is about 0.91.
This y$\rm_{cm}$ definition coincides with the rapidity in the equal velocity system for asymmetric nuclei collisions.
}
The projectile-like, target-like, and cascade-coalescence contributions of $^{3}_\Lambda$H and $^{4}_\Lambda$H are shown in different colors.
The distributions are normalized by the total number of the inelastic collisions.
Experimental data from Ref. \cite{Rappold2015} is renormalized by the rapidity bin size (See Fig. 3 of Ref. \cite{Rappold2015}).
The \emph{y}-axis therefore stands for the multiplicity per inelastic collision per unit of rapidity.
Consistent with the cross sections, the amplitudes of the rapidity distributions of the projectile-like $^{3}_\Lambda$H and $^{4}_\Lambda$H are much smaller than the data. 
We also found there exist some shift between the theoretical and experimental rapidity distributions, 
which is due to the dissipative processes in the cascade calculation. 
In addition, we found $^{3}_\Lambda$H coming from the coalescence of the cascade particles can also locate around the beam rapidity region. 
With these contributions, the theoretical yield of $^{3}_\Lambda$H still under estimate the HypHI data by more than one order of magnitude.
In the case of $^{4}_\Lambda$H, there is no contribution from cascade-coalescence particles in the beam rapidity region.
To estimate the maximum yields of $^{3}_\Lambda$H and $^{4}_\Lambda$H, we also give the rapidity distributions of 
mother-hypernuclei for $^{3}_\Lambda$H and $^{4}_\Lambda$H, which are shown by the red dash lines.
The mother-hypernuclei represent all the possible 
hot primary hyperresidues
which could decay to $^{3}_\Lambda$H or $^{4}_\Lambda$H. 
Still, it's found that even in this case the estimated maximum yields 
are smaller than the experimental data.
Nevertheless, we should mention that our calculated total yield of $\Lambda$ hyperons is around 4.5 mb,
which is even larger than the HypHI acceptance-corrected result of 1.7 $\pm$ 0.8 mb \cite{Rappold2015}.
If the reason of the low hypernuclear yields in the calculation is in a too small probability for capture of hyperons by residues in our model 
then one can consider our calculations as lower limit predictions. 
This makes future hypernuclear experiments even more promising, since it allows for a knowledge on the hyperon capture potential in excited nuclei, which, in principle, could be larger than in cold nuclei.
In panel (c) and (d), we give the transverse momentum distributions of $^{3}_\Lambda$H and $^{4}_\Lambda$H. 
The experimental data from the HypHI experiment are shown by black points. 
The spectrum in red and blue color, represents the contribution of projectile-like and target-like hypernuclear species, respectively.
In this plot, the theoretical distributions are normalized by the data.
From the transverse momentum distributions
we can clearly see that the observed $^{3}_\Lambda$H and $^{4}_\Lambda$H in the HypHI experiment 
are mainly the projectile-like component,
as the cascade-coalescence hypernuclei 
have a much broader transverse momentum distribution.

We have found that recently there was another attempt to describe
the HypHI data, 
which was published as a conference proceeding \cite{Fevre2016}. 
The authors have used the IQMD dynamical description with a coalescence-like procedure (FRIGA semi-classical method) 
for formation of clusters 
including hypernuclei. 
Within this method the hyper-clusters are considered as cold nuclei without subsequent de-excitation.
Therefore, it can be applied only for lightest species of hypernuclei. 
The calculated distributions of $^{3}_\Lambda$H and $^{4}_\Lambda$H 
over the full range of rapidity are presented.
Note that the experimental data presented in Fig.3 of Ref.\cite{Fevre2016} was not normalized by the rapidity bin size of 0.02 as presented in Fig.3 of Ref \cite{Rappold2015},
and this makes impression of low experimental yields.
With correct normalization, the yields in Ref. \cite{Fevre2016} in the projectile-like rapidity region are essentially smaller than the experimental ones, similar to our results. 
Still there are discrepancies of our calculations with the IQMD rapidity distributions of hypernuclei, 
since they predict a nearly symmetric form respective to the central rapidity. 
In our case, a heavier target (carbon) leads to enhanced production of hypermatter in the target rapidity region. 
Our difference between projectile- and target-like rapidity 
of all fragments is smaller because
of the kinetic energy loss during the particle production. 
Also relative to the experimental transverse momentum distributions of HypHI,
there is about 0.1 GeV/c shift from the calculations in Ref. \cite{Fevre2016}.
While in our case, the overall shape of the experimental transverse momentum distributions can be well described by the projectile-like component.
No shift was found in our calculation results in the case of $^{3}_\Lambda$H and a very small shift was found in the case of $^{4}_\Lambda$H.

\section{Predictions for future experiments}

In the following we focus on light hypernuclei and investigate 
the dependence of cross sections on beam energies
as well as on different projectile-target combinations. 
We restrict this study to carbon projectiles at energies of several GeV/nucleon and consider carbon and hydrogen targets.

\begin{figure}[t!]
   \setlength{\abovecaptionskip}{0pt}
   \setlength{\belowcaptionskip}{0pt}
   \begin{center}
   \includegraphics[width=8cm]{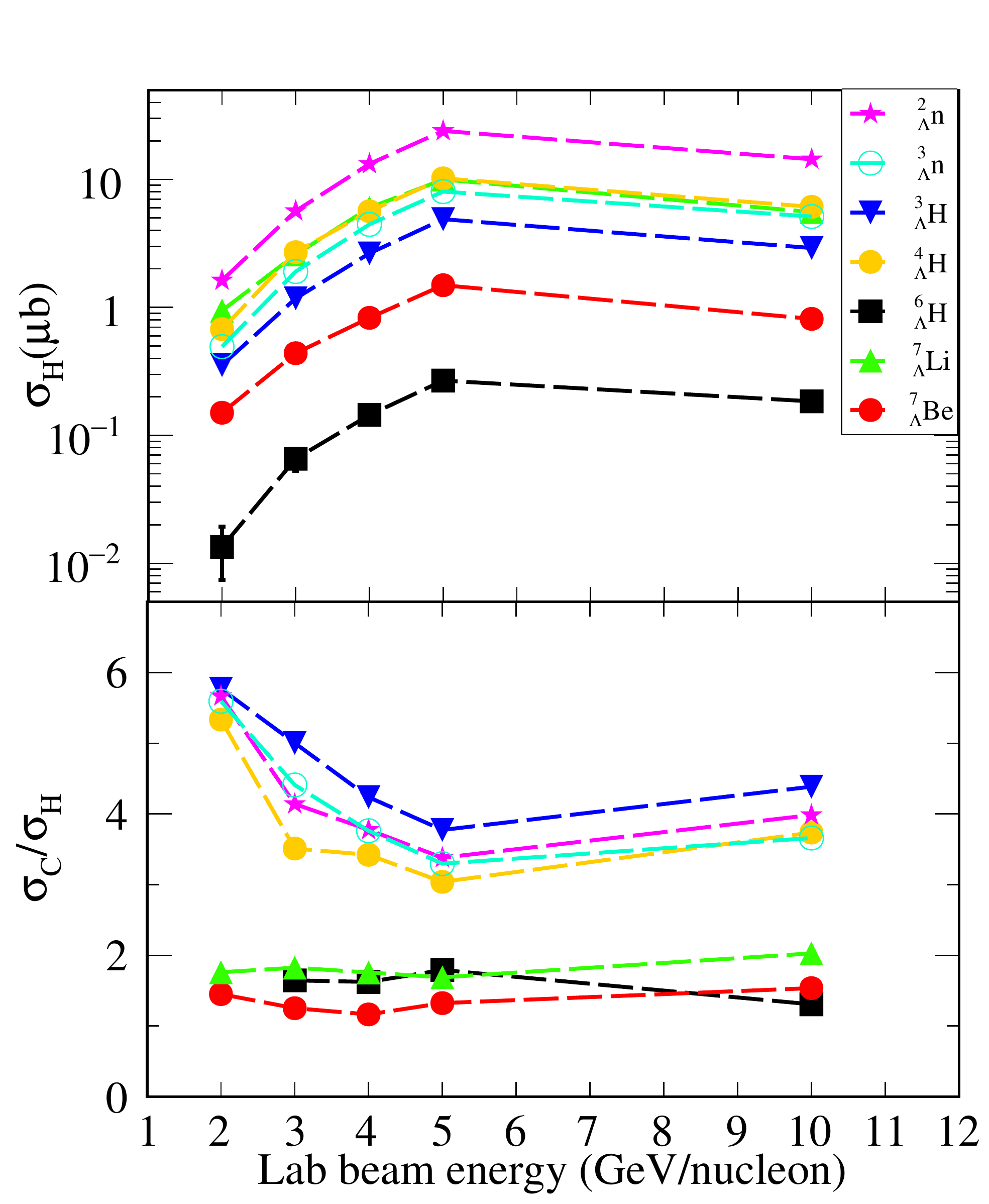}
   \end{center}
   \caption{(Color online) Top panel, energy dependence of the hypernuclear production cross section in $^{12}$C + $^{1}$H collisions at different beam energies. 
Only the statistical error is considered. 
Bottom panel, ratios of hypernuclear production cross sections between $^{12}$C + $^{12}$C and $^{12}$C + $^{1}$H collisions at different beam energies. 
Ratios with large errors are removed.
Only projectile-like hypernuclei are considered.
}
   \label{fig:cx_energy_C12H1_and_ratio}
   \end{figure}

In the top panel of Fig. \ref{fig:cx_energy_C12H1_and_ratio}
the production cross section of $^{12}$C beam impinging on a hydrogen target with energies from 1 to 10 GeV/nucleon is shown for 
$^{2,3}_{\Lambda}$\emph{n}, $^{3,4,6}_{\Lambda}$H, $^{7}_{\Lambda}$Li and $^{7}_{\Lambda}$Be hypernuclei.
The cross sections of all hypernuclei present a steep increase at the production threshold (1.6 GeV/nucleon) and a linear increase up to 5 GeV/nucleon.
A saturation plateau is shown at higher incident energies for all species. 
The calculation was also performed with carbon target, and a similar trend of the energy dependence is observed.
Such saturation behavior has also been reported in previous calculations \cite{Botvina2013, Botvina2017},
which is due to 
the balance between the amount of hyperons and residues with suitable energies that the capture may happen.
In bottom panel of Fig. \ref{fig:cx_energy_C12H1_and_ratio}, we show the ratios of hypernuclear production cross sections between $^{12}$C + $^{12}$C and $^{12}$C + $^{1}$H collisions at different beam energies.
The total inelastic cross sections for a given projectile-target combination do not vary much from 1 to 10 GeV/nucleon.
At 2 GeV/nucleon, the inelastic cross section is calculated to be 908 mb for $^{12}$C + $^{12}$C while it is 268 mb for $^{12}$C + $^{1}$H.
At 10 GeV/nucleon, the calculated inelastic cross section is 892 mb and 271 mb, respectively.
Therefore, the ratios of the total inelastic cross sections are almost a constant of 3.4 for all of the beam energies.
We notice that the cross-section ratios of hypernuclei are separated into two groups by the total inelastic cross-section ratio.
For 
light hypothetical species
$^{2,3}_{\Lambda}$\emph{n} and $^{3,4}_{\Lambda}$H, the ratios between hydrogen and carbon targets are roughly equal to the ratios of total inelastic cross sections.
With the increasing of beam energies,
there is a general saturation of the excited hyper-residual yields.
However, the carbon target favors the production of 
larger and more excited residues which decay predominantly into small fragments.
For large species $^{6}_{\Lambda}$H, $^{7}_{\Lambda}$Li and $^{7}_{\Lambda}$Be, the production cross sections between carbon and hydrogen targets are comparable through all of the beam energies.
The low ratios reflect that the corresponding fraction of low-excited large residues is smaller.
In the following discussion, we restrict the incident energies to be 2 GeV/nucleon.
Cross sections at higher energies could be
evaluated easily from Fig. 5.

   \begin{figure}[t!]
   \setlength{\abovecaptionskip}{0pt}
   \setlength{\belowcaptionskip}{0pt}
   \begin{center}
   \includegraphics[width=8cm]{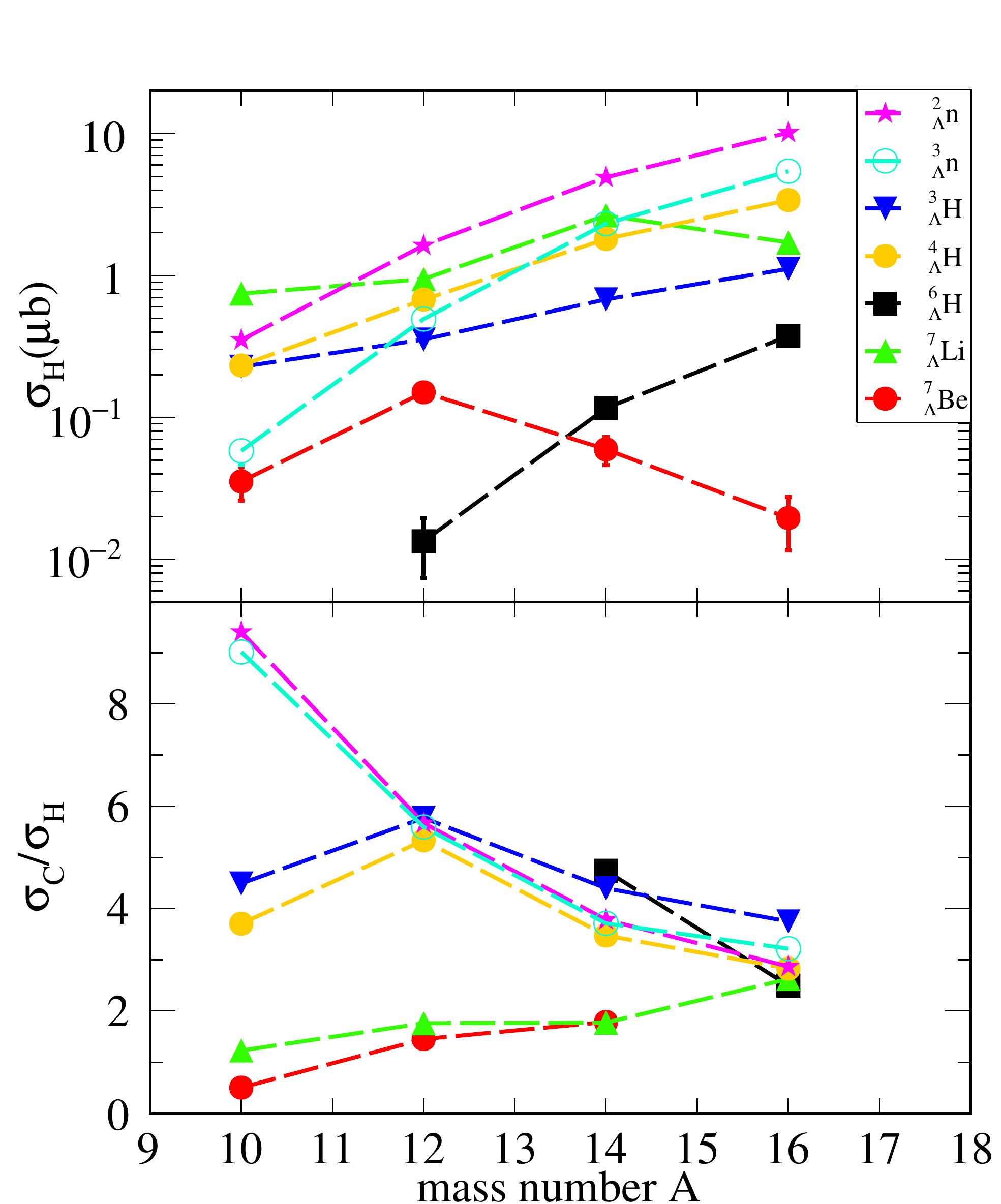}
   \end{center}
   \caption{(Color online) Top panel, projectile mass-number dependence of the hypernuclear production cross section in $^{A}$C + $^{1}$H collisions.
Only the statistical error is considered. 
   Bottom panel, ratios of hypernuclear production cross sections between $^{A}$C + $^{12}$C and $^{A}$C + $^{1}$H collisions. 
Ratios with large errors are removed.
The beam energy is fixed at 2 GeV/nucleon and only projectile-like hypernuclei are considered.
}
   \label{fig:cx_mass_C12H1_and_ratio}
   \end{figure}

We now investigate the projectile dependence of the production cross sections. 
Top panel of Fig. \ref{fig:cx_mass_C12H1_and_ratio} shows the production cross sections for 
$^{2,3}_{\Lambda}$\emph{n}, $^{3,4,6}_{\Lambda}$H, $^{7}_{\Lambda}$Li and $^{7}_{\Lambda}$Be 
for $^{10,12,14,16}$C beams impinging on hydrogen target. 
Intuitively, one could expect that neutron-rich beams may favor the production of neutron rich hypernuclei.
Indeed, as shown in top panel of Fig. \ref{fig:cx_mass_C12H1_and_ratio}, the cross sections of neutron-rich hypernuclei
($^{2,3}_{\Lambda}$\emph{n} and $^{3,4,6}_{\Lambda}$H) increase 
as much as two orders of magnitude when the projectile changes from $^{10}$C to $^{16}$C, 
while the cross sections of proton-rich hypernuclei ($^{7}_{\Lambda}$Be and $^{7}_{\Lambda}$Li) begin to decrease for $^{16}$C projectile.
The calculation with a $^{12}$C target was also performed, showing similar results. 
This indicates that the use of high-intensity neutron-rich beams may be an advantage for neutron-rich hypernuclear production.
Such measurement can be performed for secondary beam intensity around 10$^{5}$-10$^{7}$ pps since most of setups are limited by the trigger rates they can handle.
If we assume a cross section of 1 $\mu$b and a 25-cm-thick hydrogen target, 
depending on the beam intensity mentioned above,
the hypernuclear production rates will be around 0.1-10 pps,
which are high enough for a invariant-mass spectroscopy study.
The decay of the hypernuclei could happen inside of the target. 
Given the high beam energy and the small stopping power of the hydrogen target,
the use of such thick hydrogen target does not have large effects on the invariant-mass resolution.

In the bottom panel of Fig. \ref{fig:cx_mass_C12H1_and_ratio} the ratios of production cross sections between a carbon target and a hydrogen target are illustrated. 
When the projectiles are proton rich, $^{10}$C for example, 
the use of a carbon target results in 5 or 10 times larger cross sections than the use of a hydrogen target 
in the case of producing light neutron-rich hypernuclei.
This is because the yields of such nuclei on the hydrogen target are low, and charge-exchange reactions on the carbon target can increase essentially the neutron content of the projectile residue.
As the projectile's mass number increases, hydrogen target tends to have comparable production cross sections with carbon target for any hypernuclei.
The gain factor changes from 2 to 4.
Such losses should be easily compensated by using a thicker hydrogen target, since the much smaller energy loss and smaller inelastic cross sections.
The calculated inelastic reaction cross section
of $^{12}$C on hydrogen target at 2\emph{A} GeV is 268 mb while it is 908 mb for a carbon target.
With the same beam intensity, 25-cm-thick hydrogen target results in the same luminosity as 9.5-cm-thick carbon target, while the energy loss in the carbon target is more than 5 times larger.

\section{Signal-over-background ratio}

   \begin{figure*}[t!]
   \setlength{\abovecaptionskip}{0pt}
   \setlength{\belowcaptionskip}{0pt}
   \centering
   \includegraphics[width=14.5cm]{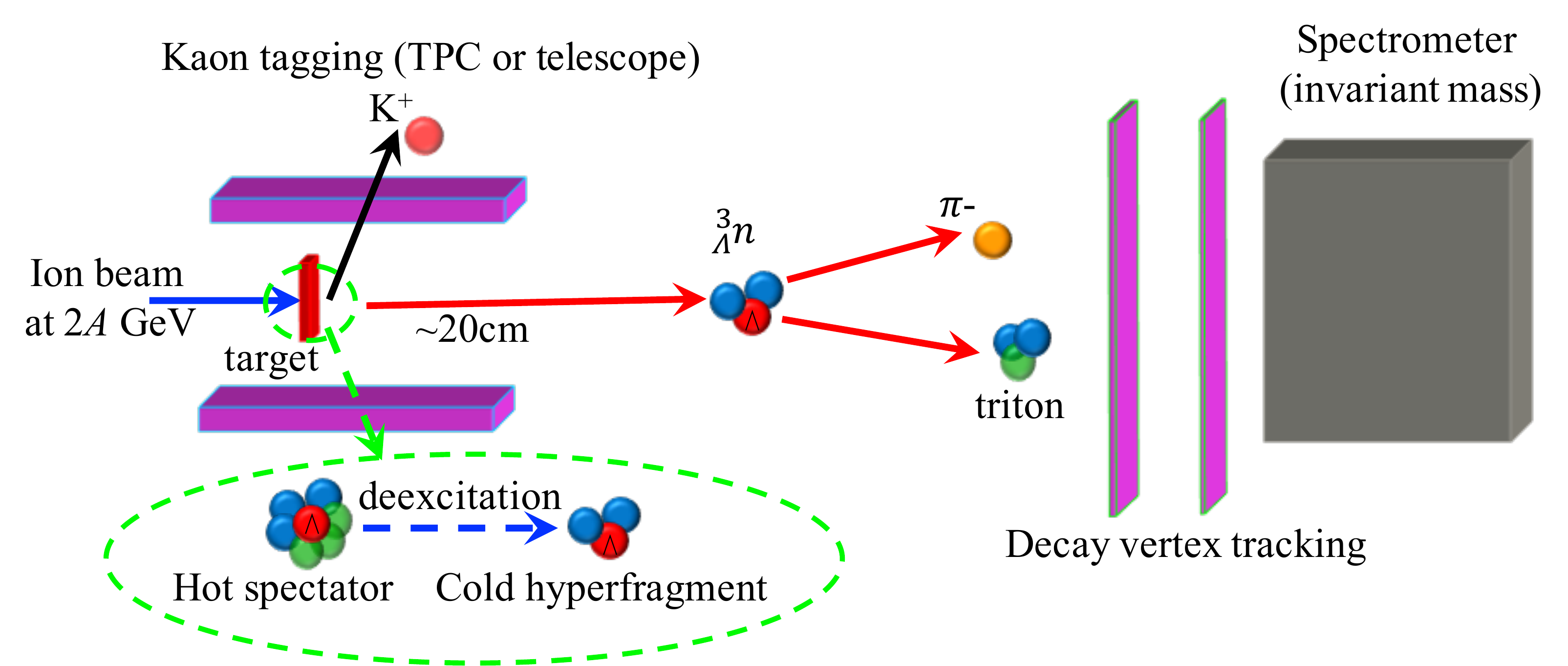}
   \caption{(Color online) Schematic diagram of the experimental setup for the invariant-mass spectroscopy of hypernuclei tagged with Kaon.}
   \label{fig:11}
   \end{figure*}

   \begin{figure}[t!]
   \setlength{\abovecaptionskip}{0pt}
   \setlength{\belowcaptionskip}{0pt}
   \centering
   \includegraphics[width=9.4cm]{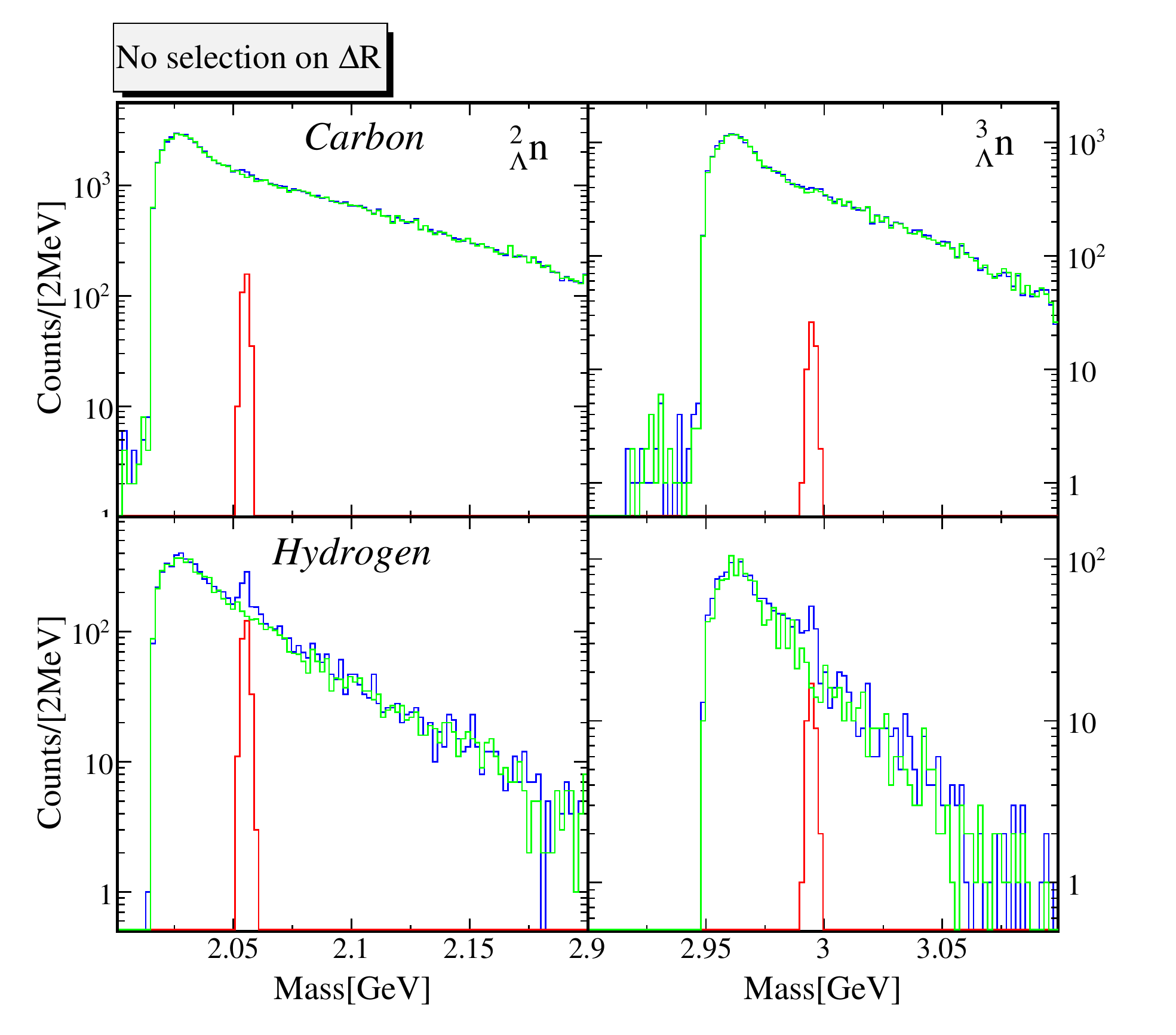}
   \includegraphics[width=9.4cm]{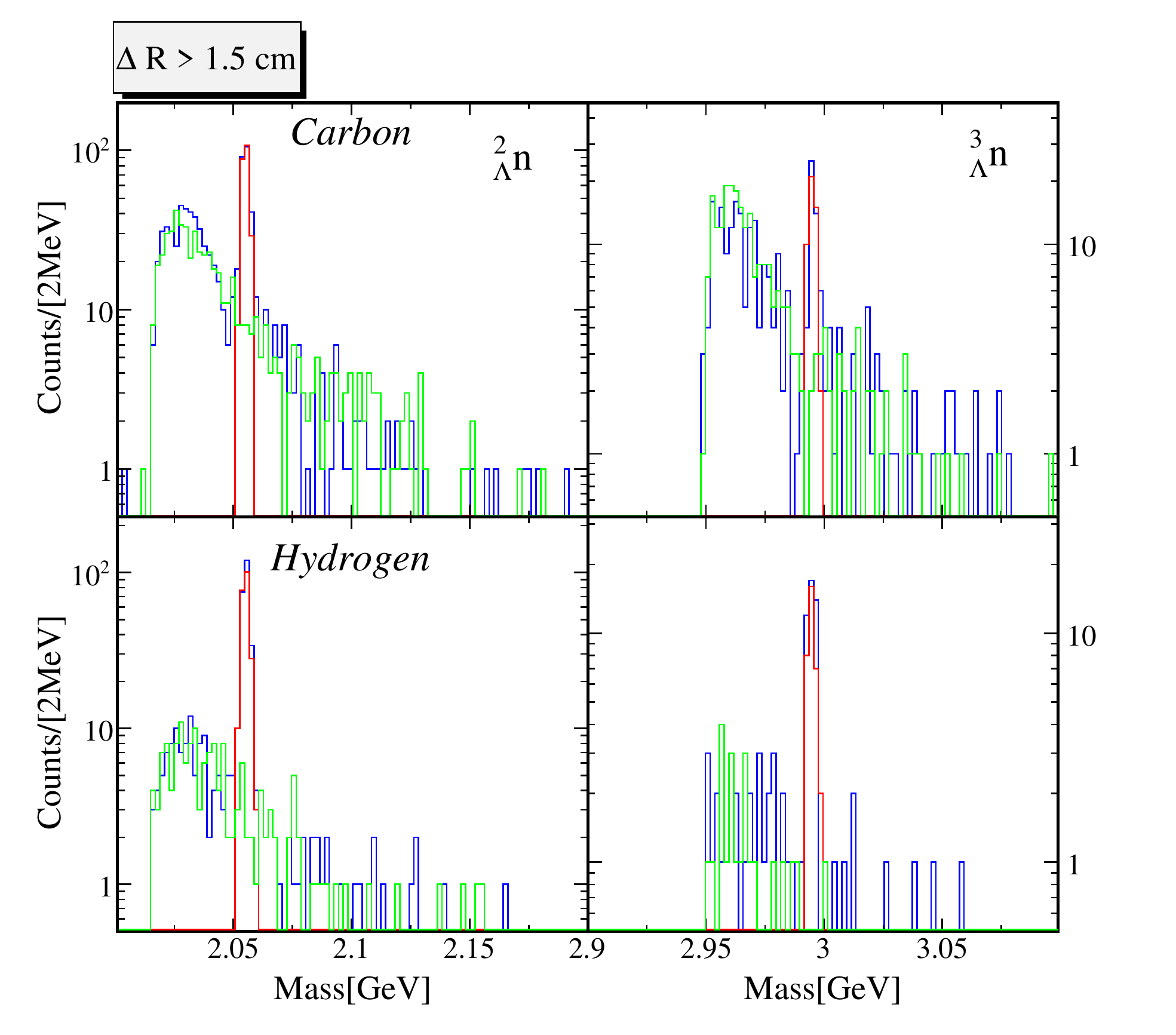}
   \caption{(Color online) 
Invariant-mass spectrums of $^{2}_\Lambda$\emph{n} and $^{3}_\Lambda$\emph{n} using $^{12}$C beam at 2 GeV/nucleon impinging on carbon target and hydrogen target. 
$\Delta$$\emph{R}$ denotes the distance between production and decay vertices.
For the figures in the top panel, there is no selection on $\Delta$$\emph{R}$ in the invariant-mass reconstruction.
For the figures in the bottom panel, $\Delta$$\emph{R}$ is selected to be larger than 1.5 cm.
The red spectrums show the invariant mass of $^{2}_{\Lambda}$\emph{n} or $^{3}_{\Lambda}$\emph{n} obtained using only the corresponding decay particles. 
The green spectrums show the background contaminations obtained using uncorrelated particles.
The blue spectrums are obtained if we consider both uncorrelated particles and decay particles in the invariant-mass reconstruction.}
   \label{fig:10}
   \end{figure}

Hypernuclei can decay through both the mesonic and the non-mesonic weak channels.
For light hypernuclei, the mesonic decay mode is favored, in which $\Lambda$ decays to $\pi$N with \emph{Q} value around 40 MeV, similar to the decay of a free $\Lambda$.
As a consequence of this decay, the $\Lambda$ is substituted by a nucleon and a pion is emitted.
The decayed final nucleon has a momentum around 100 MeV/c, much less than the Fermi momentum of 280 MeV/c.
Therefore, in medium-heavy hypernuclei, mesonic decay is suppressed by the Pauli blocking of the nucleonic medium.
In parallel,
due to a larger overlap of the wave function between $\Lambda$ and nucleons, the non-mesonic decay 
$\Lambda$N $\rightarrow$ NN
will dominate.
For non-mesonic decay,
there is no production of $\pi$ and the $\emph{Q}$ value can be as large as \emph{m}$_{\Lambda}$-\emph{m}$_{N}$ $\approx$ 176 MeV.
The final nucleons have enough energy to 
leave the nucleus or be captured and excite the nucleus.
As a result, such excited nucleus can disintegrate into nucleons and multiple heavy fragments. 
The big challenge of the invariant-mass spectroscopy is to clearly identify the production and decay of hypernuclei from their products embedded in a very high background on various particles including pions.
It is technically very difficult to measure all the non-mesonic weak decay products of medium-heavy hypernuclei.
Non-mesonic decay study of light hypernuclei such as $^{4}_{\Lambda}$He $\rightarrow$ \emph{n} + \emph{n} + \emph{p} + \emph{p}, \emph{d} + \emph{d} or \emph{p} + \emph{t} should be possible, but the branching ratios are very small \cite{Agnello2010}.
In the present invariant-mass study, we would like to 
consider
the mesonic decay of light hypernuclei.
The main contaminants in the hypernuclei identification come from the inelastic reaction channels which could result in production of $\pi$, nucleons 
and fragments without forming hypernuclei.
The decay of free $\Lambda$ and $\Sigma^{0,\pm}$ is also an important contaminant for the mesonic decay channel.
The cross sections of these processes 
may be an order of magnitude
higher than that of the cold hypernuclear production.
Intuitively, we expect that hydrogen target could provide better signal-over-background ratio than ion target, 
since the reduced number of nucleons will reduce the amount of produced pions and an exclusive measurement is in principle easier to reach.
To investigate the signal-over-background ratio quantitatively for both carbon and hydrogen target,
we perform a simulation using the theoretical events as an input.
After Fermi breakup, the produced cold hypernuclei undergo mesonic or non-mesonic decay with lifetime of hundreds of picoseconds.
Decay of $\Lambda$ and $\Sigma^{0,\pm}$ was also considered.
Due to the inverse kinematic and limitation of the acceptance, here we only focus on the projectile-like hypernuclei.
Target-like particles were removed from the simulation since they can be easily rejected by a proper momentum acceptance setting in the experiment.
Only particles with rapidity larger than zero and moving in very forward angle ($\theta\rm_{lab}$ $\textless$ 10$^\circ$) were considered as accepted. 
The reconstructed hypernuclei were required to have rapidity of y \textgreater 0.75y$\rm_{proj}$ and the scattering angles were required to be less than 5$^\circ$,
where y$\rm_{proj}$ is the rapidity of the projectile in the laboratory frame.
In our simulation, a constant momentum resolution ($\sigma_{p}$/\emph{p} = 1\%) was taken for all kinetic energies of $\pi$, proton, neutron and heavy fragments.
This leads to a resolution (FWHM) of 2.5 MeV for $\Lambda$ ($\pi$$^{-}$ + proton).
In addition, 5 mm (1$\sigma$) spatial resolution was considered for the production and decay vertices in \emph{x}, \emph{y} and \emph{z} direction.
To reduce the huge background, we require only events with strangeness production, which corresponds to coincide with kaon production 
around the target in the experiment.
The lifetime of K$^{+}$ meson is 12 ns and it will decay to $\mu^{+}$ + $\nu_{\mu}$ or $\pi^{+}$ + $\pi^{0}$ with a branching ratio of 63.5\% and 21.2\% respectively. 
It has been shown K$^{+}$ can be efficiently identified either in flight with a time projection chamber (TPC) \cite{Adam2015} or at rest using a kaon range telescope \cite{Armstrong1990}.
A schematic diagram of the experimental setup is shown in Fig. 7.

Recently, exotic bound hypernuclei, like $^2_{\Lambda}$\emph{n} and $^3_{\Lambda}$\emph{n}, were extensively discussed and looked for in relativistic ion experiments \cite{Rappold2013PRC, Donigus2015}.
As examples, we consider here
the mesonic decay processes of $^2_{\Lambda}$\emph{n} and $^3_{\Lambda}$\emph{n}, $\emph{i.e.}$ 
$^2_{\Lambda}$\emph{n} $\rightarrow$ $\pi$$^{-}$ + $\emph{d}$, 
$^3_{\Lambda}$\emph{n} $\rightarrow$ $\pi$$^{-}$ + $\emph{t}$.
Since the lifetimes of $^2_{\Lambda}$\emph{n} and $^3_{\Lambda}$\emph{n} are still unknown,
the possible lifetimes of 181 ps and 190 ps were used in the simulation for 
$^2_{\Lambda}$\emph{n} and $^3_{\Lambda}$\emph{n} \cite{Rappold2013PRC}.
Invariant masses of $^2_{\Lambda}$\emph{n} and $^3_{\Lambda}$\emph{n} were reconstructed from the momentum of $\pi$$^{-}$ + $\emph{d}$ and $\pi$$^{-}$ + $\emph{t}$, respectively.
The obtained invariant-mass spectrums are shown in Fig. \ref{fig:10}
for $^{12}$C beams impinging on carbon and hydrogen target at 2$\emph{A}$ GeV.
We note that all of the combinations were considered in the invariant-mass reconstruction 
if there were multiple $\pi$$^{-}$, $\emph{d}$ and $\emph{t}$ accepted.
For direct comparison, the number of the collision events, the experimental acceptance as well as the reconstruction of hypernuclei were the same for each plot.
From Fig. \ref{fig:10}, we can clearly see the improvement of signal-over-background ratio when using a hydrogen target.
We found that one main reason is the reduction of $\pi^{-}$ background at forward angles.
In the $^{12}$C + $^{12}$C collisions, 78\% of the $\pi^{-}$ background comes from the cascade collisions, while in the case of $^{12}$C + proton, this ratio drops to 27 \%. 
After coincidence with kaon, most of the $\pi^{-}$ background in $^{12}$C + proton collisions comes from the decay of free lambda particles, which is the main background for the mesonic decay channels.
We note that 
vertex methods for identification of products of slow weak decays have been successfully achieved in several hypernuclear experiments \cite{Alice2013, Rappold2013NPA, Rappold2013PRC}. 
From the simulation results in Ref. \cite{Minami2008}, the $\pi^{-}$ background can be reduced to 1.7\% by applying a vertex trigger in the $^{6}$Li and $^{12}$C collisions at 2\emph{A} GeV.
The background suppression could be clearly seen in the bottom panel of Fig. \ref{fig:10}, 
where the distance between the production and decay vertices ($\Delta$$\emph{R}$) was required to be larger than
1.5 cm, which is about two times of the resolution of the distance.
In our simulation, multiple scattering of pions and light-ions in the target was not considered, 
so we expect a worse signal-over-background ratio especially for carbon target.
Thus, we foresee in a future work to further study the performances of a realistic setup dedicated to hypernuclei production from hydrogen induced reactions.

\section{Summary}
Ion beam induced reactions are a very promising way to produce exotic hypernuclei,
as already proved by the HypHI collaboration at GSI.
In this article, we present a series of calculations using the Dubna intranuclear Cascade Model followed by Fermi breakup to investigate theoretically the production of light $\Lambda$ hypernuclei.
The calculated cross sections are compared with available experimental data. 
We found the Dubna data could be fairly well reproduced if we slightly tune the excitation-energy distribution of the hot primary hyperresidues.
However, the calculated yields of hypernuclei are more than one order of magnitude smaller than the recently published HypHI data.
With a more detailed comparison of rapidity and transverse momentum distributions, 
we confirm
that the observed hypernuclei in the HypHI experiment are mainly projectile-like hypernuclei with a small cascade-coalescence contribution.
Although the amplitudes are much smaller and there exist some rapidity shift, the overall shape of both rapidity and transverse momentum distributions agree with the published data.
Furthermore, we also investigate the cross-section dependence on beam energies and different projectile-target combinations.
Comparing with carbon target, hydrogen target also leads to sizable hypernuclear yields, even for exotic species.
In the presented calculations,
the cross-section ratios between carbon and hydrogen targets are similar with the total inelastic cross-section ratios, 
making hydrogen a competitive target for hypernuclear production in relativistic ion collisions.
The typical hypernuclear production cross sections at 2\emph{A} GeV beam energy with hydrogen target are around 0.5 $\mu$b.
From the experimental point of view, 
we also investigate the signal-over-background ratio using $^{12}$C beam impinging on hydrogen and carbon targets.
Invariant-mass spectrums of $^2_{\Lambda}$\emph{n} and $^3_{\Lambda}$\emph{n} are given taking into account the experimental acceptance and resolution.
With these examples, we 
demonstrate that a hydrogen target could indeed 
reduce significantly
the background contamination in the mesonic decay channel for some experiments.
Hypernuclear production data from ion collisions with hydrogen and carbon targets are required to benchmark the current predictions 
and allow for the development of future experimental programs 
at FAIR facility in GSI and HIAF facility in China.

\begin{acknowledgments}
Y. L. S. acknowledges the support of Marie Sk\l{}odowska-Curie Individual Fellowship (H2020-MSCA-IF-2015-705023) from the European Union.
A. S. B. acknowledges the support of BMBF (Germany).
The authors thank K. K. Gudima for clarifying discussions of the calculations, and J. Pochodzalla for stimulating this work.
The valuable discussions with C. Rappold, T. Saito 
and C. Scheidenberger are gratefully acknowledged.
Support from ESNT for the organizing of hypernuclear workshop at CEA is greatly acknowledged.

\end{acknowledgments}

\section*{References}

\bibliography{mybibfile}

\end{document}